\documentclass[journal,article,accept,moreauthors,dvipdfm,12pt,a4paper]{mdpi} 

\lastpage{x}
\doinum{10.3390/------}
\pubvolume{xx}
\pubyear{2013}
\history{Received: xx / Accepted: xx / Published: xx}

\usepackage{amssymb}
\usepackage{graphicx}

\Title{Folding Kinetics of Riboswitch Transcriptional Terminators and Sequesterers}

\Author{Ben Sauerwine $^{1,}$*, Michael Widom $^{1}$}

\address[1]{%
$^{1}$ Department of Physics, Carnegie Mellon University, 5000 Forbes Avenue, Pittsburgh, PA  USA  15213}

\corres{widom@andrew.cmu.edu, 412-268-7645}

\abstract{To function as gene regulatory elements in response to environmental signals, riboswitches must adopt specific secondary structures on appropriate time scales. We employ kinetic Monte Carlo simulation to model the time-dependent folding during transcription of TPP riboswitch expression platforms.  According to our simulations, riboswitch transcriptional terminators, which must adopt a specific hairpin configuration by the time they have been transcribed, fold with higher efficiency than Shine-Dalgarno sequesterers, whose proper structure is required only at the time of ribosomal binding.  Our findings suggest both that riboswitch transcriptional terminator sequences have been naturally selected for high folding efficiency, and that sequesterers can maintain their function even in the presence of significant misfolding.}


\begin{document}


\section{Introduction}
\label{sec:introduction}

The riboswitch is a mechanism of self-regulation in messenger RNA that is found primarily in metabolic genes of bacteria.
Riboswitches possess an {\em aptamer} that is capable of binding a specific ligand, and an {\em expression platform} that regulates the gene's expression according to the binding state of the aptamer~\cite{Breaker05}.
The expression platform can regulate expression through formation of an intrinsic (rho-independent) terminator hairpin, by sequestering a Shine-Dalgarno ribosomal binding site, or by cleaving the messenger.
The terminator hairpin operates by halting transcription while the Shine-Dalgarno sequesterer, also a hairpin, operates by preventing translation.

\begin{figure}
   \begin{center}
	\includegraphics[clip,width=0.75\textwidth]{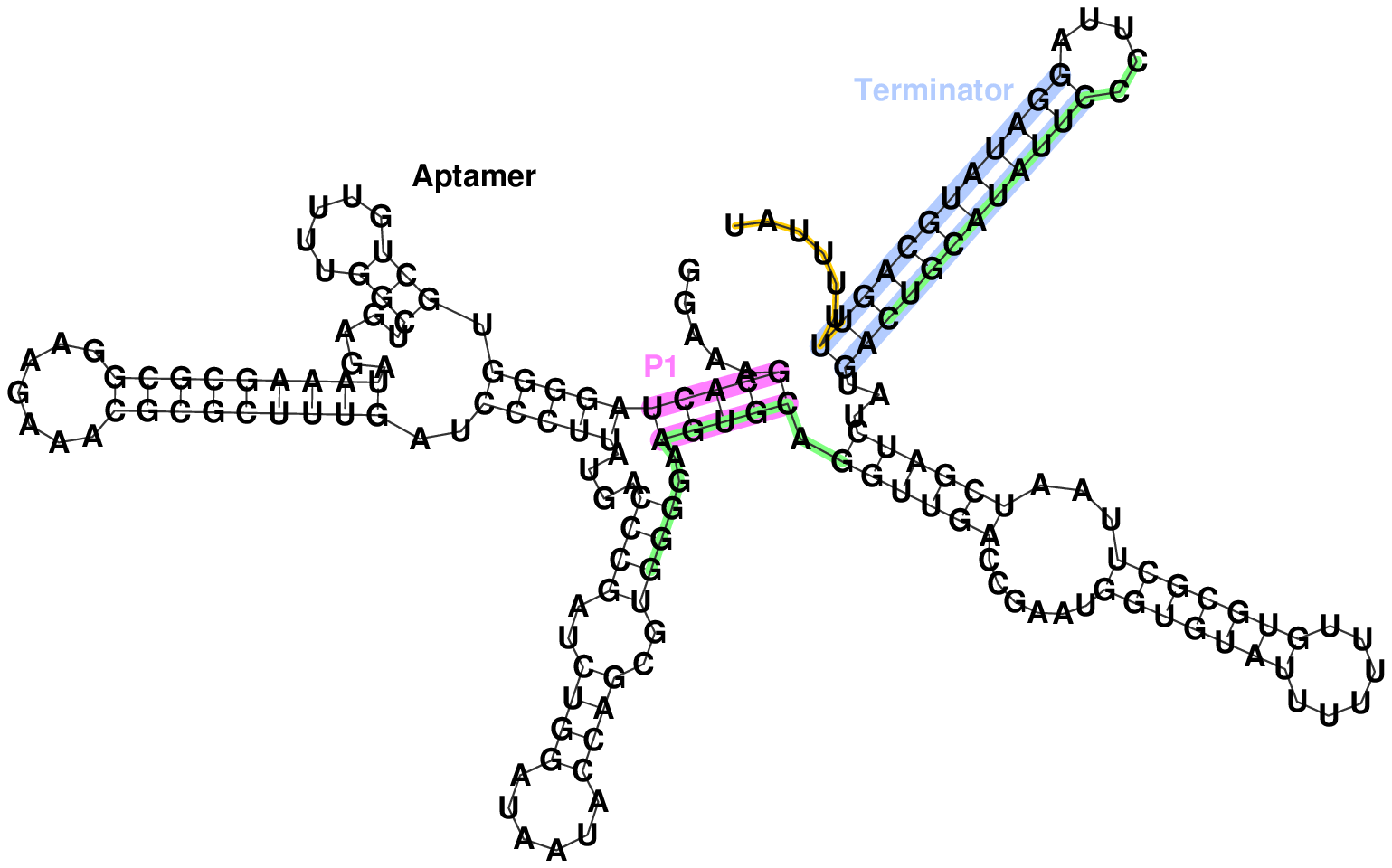}
	\includegraphics[clip,width=0.95\textwidth]{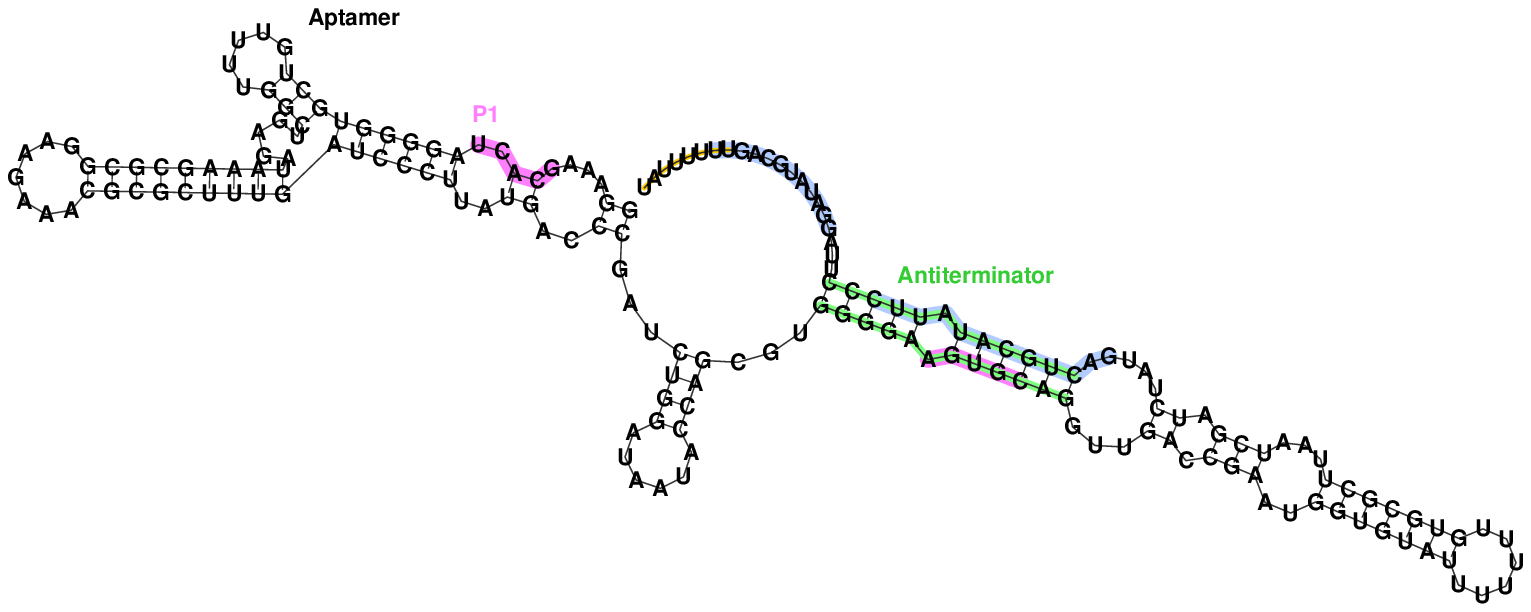}
   \end{center}
\caption{\label{fig:ykoF}
Secondary structure of the aptamer and terminator of the {\it Bacillus subtilis} ykoF riboswitch. (top) Bound state, aptamer formed, transcription off. The P1 stem of the aptamer (pink) conflicts with the antiterminator (green), allowing formation of the terminator (blue), and thus halting transcription via the poly-$U$ pause site (orange).  (bottom) Unbound state, aptamer unformed, transcription on. Destabilizing the aptamer allows formation of the antiterminator, which conflicts with the terminator, and hence allows transcription to proceed.}
\end{figure}

Because riboswitches function through conformational changes resulting from the ligand-bound or unbound state of the aptamer, they rely on both RNA thermodynamics and structural kinetics.
Of particular importance is the secondary structure, the pattern of pairing among complementary bases (see Fig.~\ref{fig:ykoF}).  
This secondary structure forms both during and after mRNA transcription~\cite{Pan06,Proctor13}, leading to a time-dependent free energy landscape for RNA folding.  
The importance of kinetics for the operation of some terminator-type riboswitches is supported by the presence of transcriptional pause sites following the aptamer and antiterminator~\cite{Wickiser05}.

Here we concentrate on the dynamical folding of the expression platform as it grows during transcription.  A minimum free energy (MFE) structure adopted by an incomplete sequence may become metastable once the sequence is complete.
The lifetime~\cite{Sauerwine11} of such a metastable structure may exceed the time allowed for the switch to function, leading to a failure of gene regulation.
By comparing the folding efficiencies of transcriptional terminator-type riboswitch terminator hairpins with that of with translational sequesterers, we suggest that riboswitch transcriptional terminators have been naturally selected to fold reliably under the time constraint imposed by the mRNA transcription rate.
By inspection of specific poorly folding sequesterers, we propose that even misfolded sequesterers may retain some function provided their Shine-Dalgarno sequence remains bound.

\section{Methods}
\label{sec:methods}

\subsection{Sequences}
\label{sec:sequences}

Riboswitch aptamers are highly conserved and well annotated in the RFam database~\cite{Rfam}.
Unfortunately, the expression platform sequences are poorly conserved and are not generally annotated.
Their hairpin topology is their main conserved feature, along with either a trailing poly-U pause site in a terminator (see Fig. 1) or a Shine-Dalgarno ribosomal binding site in a sequesterer.
We choose to study a set of TPP (thiamine pyrophosphate, vitamin B$_1$) riboswitches whose expression platforms have been independently annotated~\cite{Rodionov02}.
Out of the 135 annotated riboswitches, 73 are classified as sequester-type, 52 as terminator-type, 9 as both terminator and sequesterer, and one as neither.
We choose to examine only those expression platforms with a definite classification, and we include an additional four nucleotides on each end to provide genomic context for our folding studies.
All sequences studied fold to a hairpin as their MFE structure, as annotated.
Some statistical properties of the resulting sequences are shown in table~\ref{tab:stats}.
Notice that on average the sequesterers are longer than terminators by 10 nucleotides, and also that their lengths are highly variable.
To explore the dependence of folding efficiency on length, we constructed an artificial family of extended terminators by adding 5 nucleotide pairs randomly to each terminator, drawing the additional pairs from the pairs already present in the original terminators.
We then randomly shuffle the pairs while preserving the topology of bulges and loop in the minimum free energy structure.

\begin{table}
\begin{center}
\begin{tabular}{r|ccc}
                     & Sequesterers   & Terminators  & Extended \\
\hline
Length (nucleotides) & 47.7$\pm$11.2  & 37.8$\pm$5.7 & 47.8$\pm$5.7 \\
MFE (kCal/mol)       & -14.9          & -16.3        & -27.4 \\
MFE frequency        & 0.35           & 0.52         & 0.53 \\
Ensemble diversity   & 1.97           & 0.70         & 0.58 \\
\hline
A \%                 & 23.8           & 21.7         & 22.3 \\
C \%                 & 26.1           & 18.3         & 18.5 \\
G \%                 & 23.7           & 19.7         & 20.9 \\
U \%                 & 26.4           & 40.3         & 38.3 \\
\end{tabular}
\end{center}
\caption{\label{tab:stats} Sequesterer, terminator and extended terminator sequence average properties.  MFE frequency and ensemble diversity represent the frequency of the MFE structure and the diversity of the secondary structure ensemble at T=37$^\circ$C, as obtained from {\tt RNAfold}~\cite{ViennaRNA}. A, C, G and U are nucleotide fractions.}
\end{table}

\subsection{Folding}
\label{sec:folding}

For Minimum Free Energy calculations of secondary structures, {\tt RNAfold}~\cite{ViennaRNA} is used with the {\tt ViennaRNA} 1.4 energy model~\cite{Mathews99} at temperature $T=37^{\circ}$C.
Note that we will not be able to capture the influence of tertiary contacts or of pseudoknots within the confines of this model.
The default energy parameters for $1$M NaCl are used despite cellular conditions being $150-250$ mM Na$^{+}$ and $5-10$ mM Mg$^{2+}$, since the energetics of the secondary structures in these conditions are similar~\cite{Tan07}.
That is, 1M NaCl has approximately equivalent ionic strength to real cellular conditions, since the doubly-charged Mg$^{2+}$ is far more effective at compensating the phosphate backbone of nucleic acids than the singly charged Na$^{+}$.
A suitable validated energy model for true cellular conditions is not available~\cite{Liu05}.

Folding is simulated at the level of secondary structure by kinetic Monte Carlo using the ViennaRNA program {\tt kinfold}~\cite{Flamm00}.
The rate for transitions in {\tt kinfold} is given in arbitrary units that require calibration to real time.
As an estimate for the {\tt kinfold} timescale, $\tau_{K}$ is taken to be about $5 \mu s/{\rm step}$, from the calibration of Liu and Ou-Yang~\cite{Liu05}.
To simulate folding during transcriptional growth, additional nucleotides are added to the $3^{\prime}$ end of the chain at regular time intervals.
Typical bacterial transcription rates, $R_t$, range from 20-80 nt/s, with 50 nt/s taken as standard.
The possibility that a significant transcriptional pause might take place inside the antiterminator or the terminator is neglected though this could be important in some specific cases~\cite{Wickiser05}.
Simulating at the level of secondary structure is more efficient, though less realistic, than applying molecular dynamics to coarse-grained continuum models~\cite{Ding2008,Denesyuk2013}.

Because we focus on the competition between folding rates and transcription rates, the chief parameter governing the simulation is the product $\rho=1/(\tau_K R_t)$, representing Monte Carlo step performed between each nucleotide addition.  Our standard value is $\rho=4000$ MC steps/nt transcribed.  Because of the range of transcription rates $R_t$, as well as the uncertainty concerning the timescale calibration $\tau_K$, we carry out simulations over a range of values of $\rho$.  Our primary result, the high efficiency of terminator folding relative to sequesterers, holds over several orders of magnitude in $\rho$.

\subsection{Statistical analysis of distributions}
\label{sec:stats}

\begin{figure}
\begin{center}
\includegraphics[clip,,angle=-90,width=0.99\textwidth]{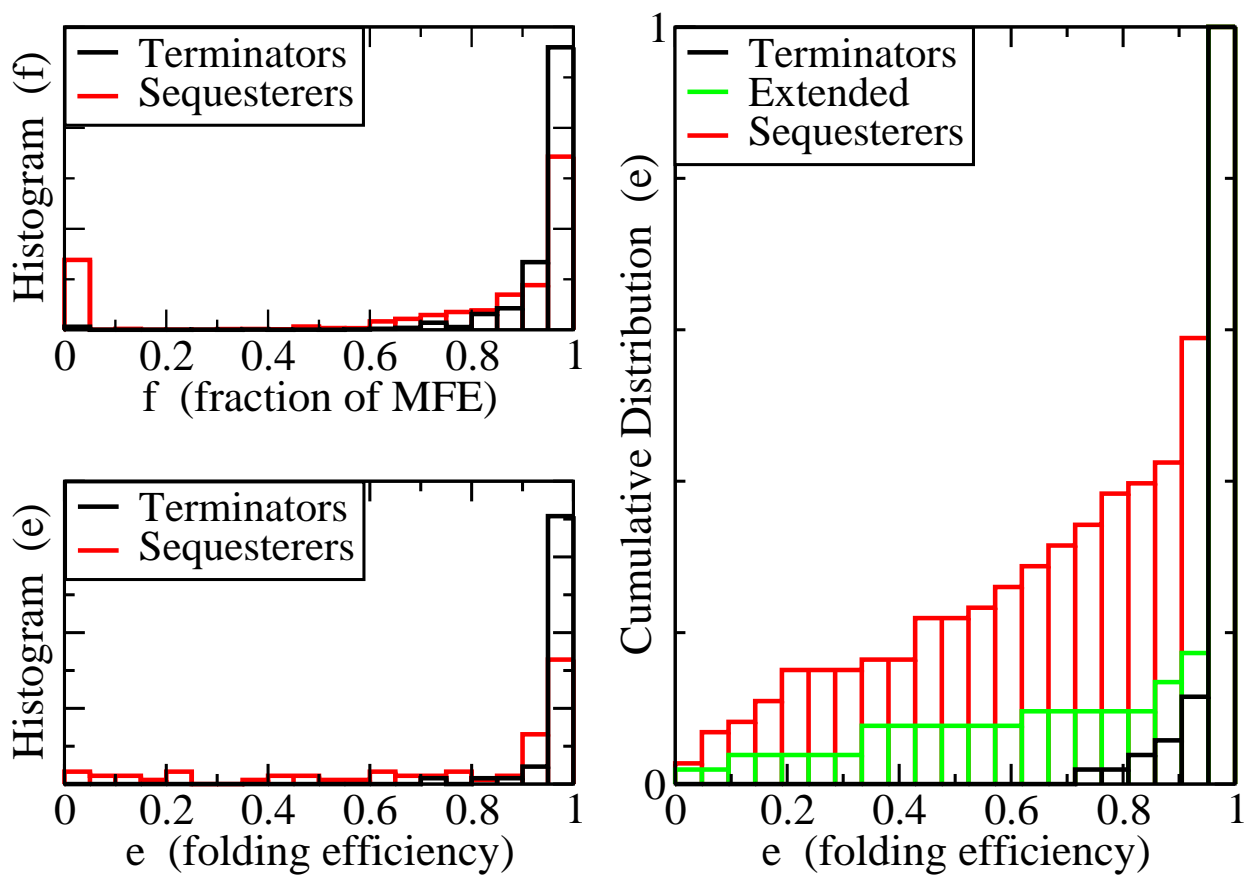}
\end{center}
\caption{\label{fig:eff}
Normalized distributions of folding performances for terminator- and sequesterer-type riboswitches from a wide range of prokaryotes.
(a) Histograms of folding fractions $f$ combined for all sequences $s$ of each specific type.
(b) Histograms of folding efficiencies $e_s$ for individual sequences $s$.
(c) Cumulative distribution of folding efficiencies including extended terminators.}
\end{figure}

Results of this study will be presented in the form of distributions over repeated kinetic folding attempts for many individual sequences.
For example, Fig.~\ref{fig:eff}a displays a histogram showing the relative frequency $P(f)$ with which terminators reach a given fraction $f$ of their MFE structures under our standard growth conditions.
According to this figure, in our complete population of transcriptional terminators, each folded 100 times, 100\% of the expected nt pairs are obtained in the majority of trials, and more than 60\% of the expected pairs are obtained in all the trials.
However, 0\% of the expected pairs are obtained in a non-negligible subset of trials for Shine-Dalgarno sequesterers.
It is not known what fraction of the MFE hairpin structure is required for successful termination, but one can set a threshold $t$ anywhere between 1\% and 80\% with almost no impact on the fractions $f$ of terminator folds that lie above and below this threshold, because the distribution nearly vanishes over this range.
We use the fraction of expected pairs as a metric for termination, rather than the free energy of the folded structure, because deep metastable traps are precisely what is to be avoided for efficient folding and termination.

\begin{figure}
\begin{center}
\includegraphics[clip,angle=-90,width=0.99\textwidth]{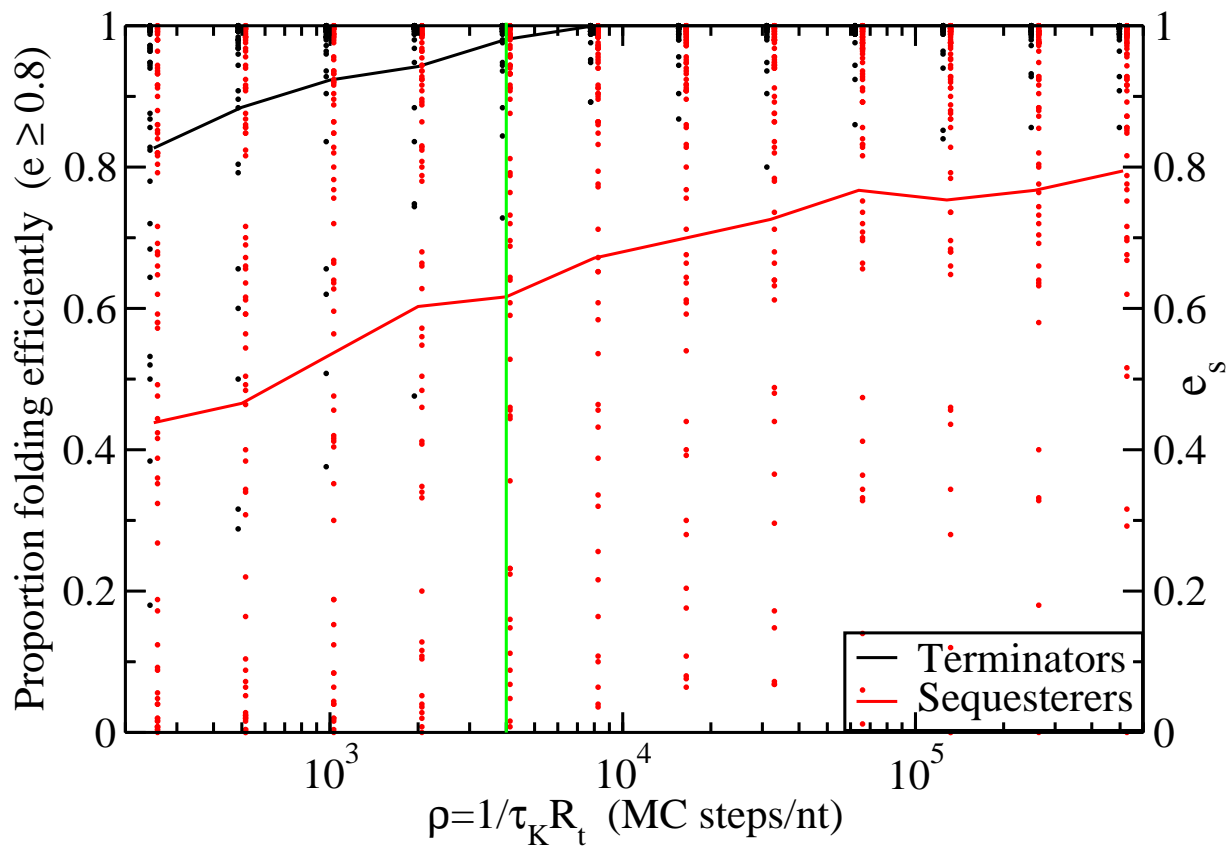}
\end{center}
\caption{\label{fig:efftime}
Proportion of TPP terminators (black line) and sequesterers (red line) that fold efficiently (i.e. with $e\ge 0.8$) at various timescales $\rho$.
Data points indicate the individual folding efficiencies $e_s$ of each hairpin sequence $s$.
Green line at $\rho=4000$ MC steps/nt transcribed indicates the timescale for $\tau_{K} = 5 \mu$s and $R_{t} = 50$ nt/s.}
\end{figure}

For a given sequence $s$, its folding efficiency $e_s$ is defined as the fraction of attempted folds that form a viable hairpin.
Define the viability $v$ of a fold as a function of the MFE structure fraction $f$ through the equation $v(f)=\theta(f-t)$, where $\theta$ is the step function and $t$ is the viability threshold.
Thus $e_s=\langle v(f)\rangle$ averaged over independent folding attempts.
If sequence $s$ has probability distribution $P_s(f)$ for folding to MFE fraction $f$, its efficiency can be evaluated as $e_s=\int_0^1P_s(f)v(f){\rm d}f=\int_t^1P_s(f){\rm d}f$.
This is precisely the fraction of attempted folds whose folding fraction $f$ lies above the threshold $t$.
As discussed above, considerable freedom exists in the choice of the threshold $t$, but $t=0.7$ is taken as a reasonably conservative limit because a high fraction of the MFE structure is presumably required for actual functionality of the terminator.  Hence we define a structure with a fraction $f$ of its MFE base pairs below $t=0.7$ as ``misfolded''.

\section{Terminators vs. sequesterers}
\label{sec:results}

Riboswitch intrinsic terminator hairpins can be expected to fold with greater efficiencies than sequesterers because terminators act at the time of transcription.  The constraint that terminators must perform within the transcription time means that terminators must fold quickly.  Meanwhile, sequesterers act at the time of translation, effectively relaxing this constraint.
Here the folding efficiencies of the sequesterers are compared to transcriptional terminator hairpins across a family of riboswitches.
TPP-binding riboswitches are chosen, because of the availability of annotated terminator and sequesterer riboswitches~\cite{Rodionov02}.

According to Fig.~\ref{fig:eff}, the terminator hairpins do indeed fold quite efficiently, with all but one of Rodionov's annotated terminators having a folding efficiency greater than 80\% under our standard growth conditions.
However, the sequesterers fold with substantially lower efficiency.
Table~\ref{tab:chisq} enumerates the numbers of hairpins of each type folding efficiently ($e\ge 80\%$) and inefficiently ($e<80\%$).
The $p$-value for the null hypothesis ({\em i.e.} the assertion that the proportion of efficient sequesterers equals the proportion of efficient terminators) is $p=3\times 10^{-7}$ (Fisher exact test), providing strong support for the claim that terminator-type riboswitch hairpins fold with higher efficiency during transcription than do sequesterer-types.
Figure~\ref{fig:efftime} shows the proportion of efficiently folding terminators and sequesterers for a range of timescales $\rho=1/(\tau_K R_t)$, allowing for $\tau_{K}$ and $R_{t}$ to vary over orders of magnitude without affecting the conclusion that terminators fold with higher efficiency than sequesterers.

\begin{table}[h!]
  \begin{center}
    \begin{tabular}{| l | l | l |}
    \hline
                & Efficient   & Inefficient \\
                &($e \ge 0.8$)&($e < 0.8$)\\
    \hline
    Terminator  & 51          & 1 \\
    Sequesterer & 45          & 28 \\
    \hline
    \end{tabular}
  \end{center}
\caption{\label{tab:chisq} Efficiency table for Fisher's exact test comparing terminator hairpins to Shine-Dalgarno sequesterers, when grown at 50 nt/s and assuming $\tau_{K} = 5 \mu s$.}
\end{table}

What explains the relative folding efficiencies of terminators and sequesterers?
As outlined in Table~\ref{tab:stats}, some gross features of rho-independent terminator sequences differ from Shine-Dalgarno sequesterers.
Perhaps the primary difference between them lies in their length distributions.
TPP rho-independent terminators are 38 nucleotides long on average, while the Shine-Dalgarno sequesterers average 48 nucleotides in length.
Indeed, longer sequences will tend to possess more and deeper metastable states that would compete with the MFE state.
The length dependence of folding efficiency was tested by duplicating 5 base pairs in each TPP rho-independent terminator in order to mimic the lengths of sequesterers.
The results shown in Fig.~\ref{fig:eff}c, indicate that while there is some effect detrimental to efficient folding in the longer hairpins, this length difference alone does not account for the difference in folding efficiencies.
Furthermore, while we note a weak correlation of decreasing efficiency with increasing terminator sequence length, neither the extended terminators nor the sequesterers exhibit any significant correlation between efficiency and length.

\begin{figure}
   \begin{center}
	\includegraphics[clip,width=0.35\textwidth]{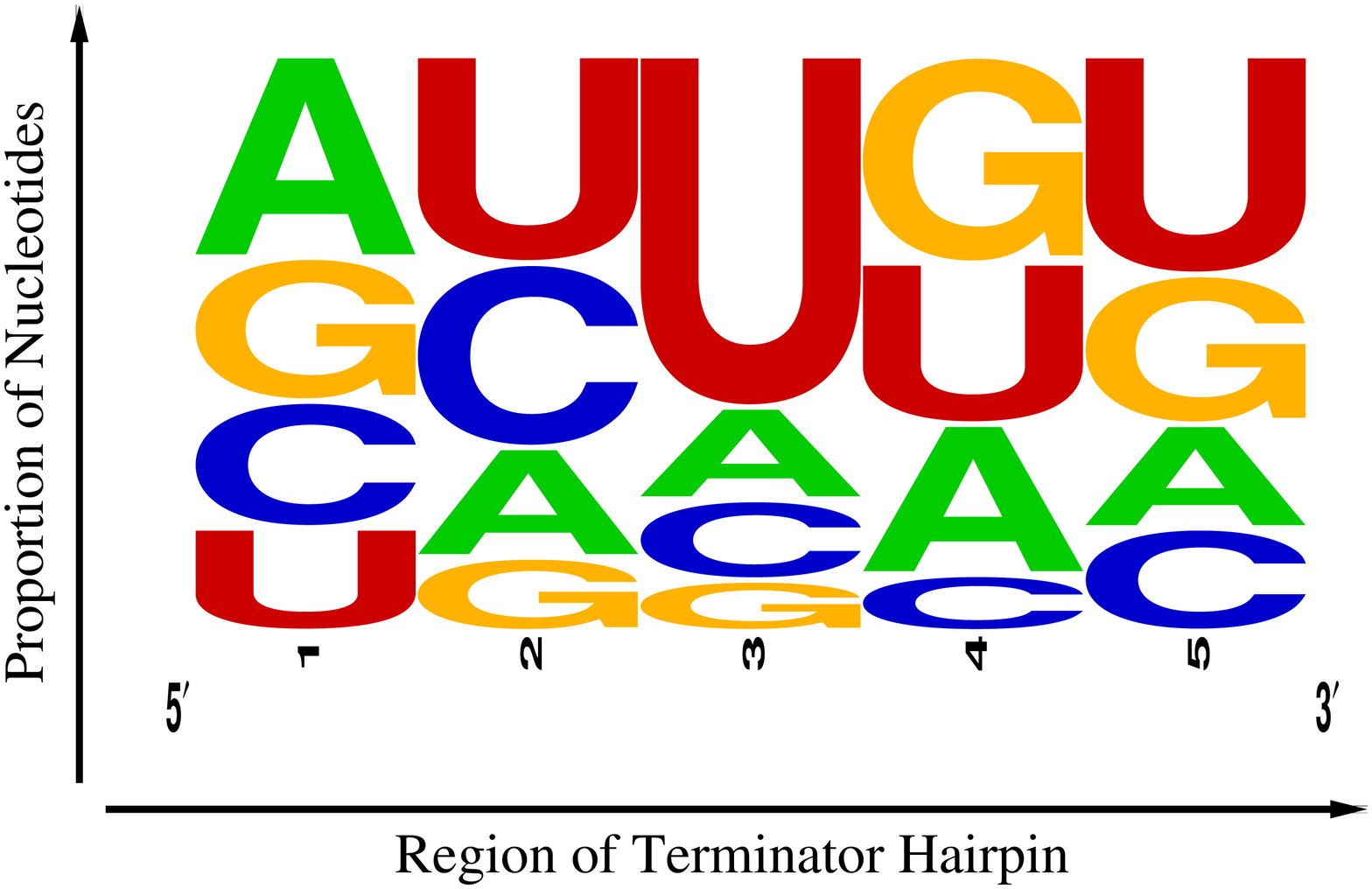}
	\includegraphics[clip,width=0.35\textwidth]{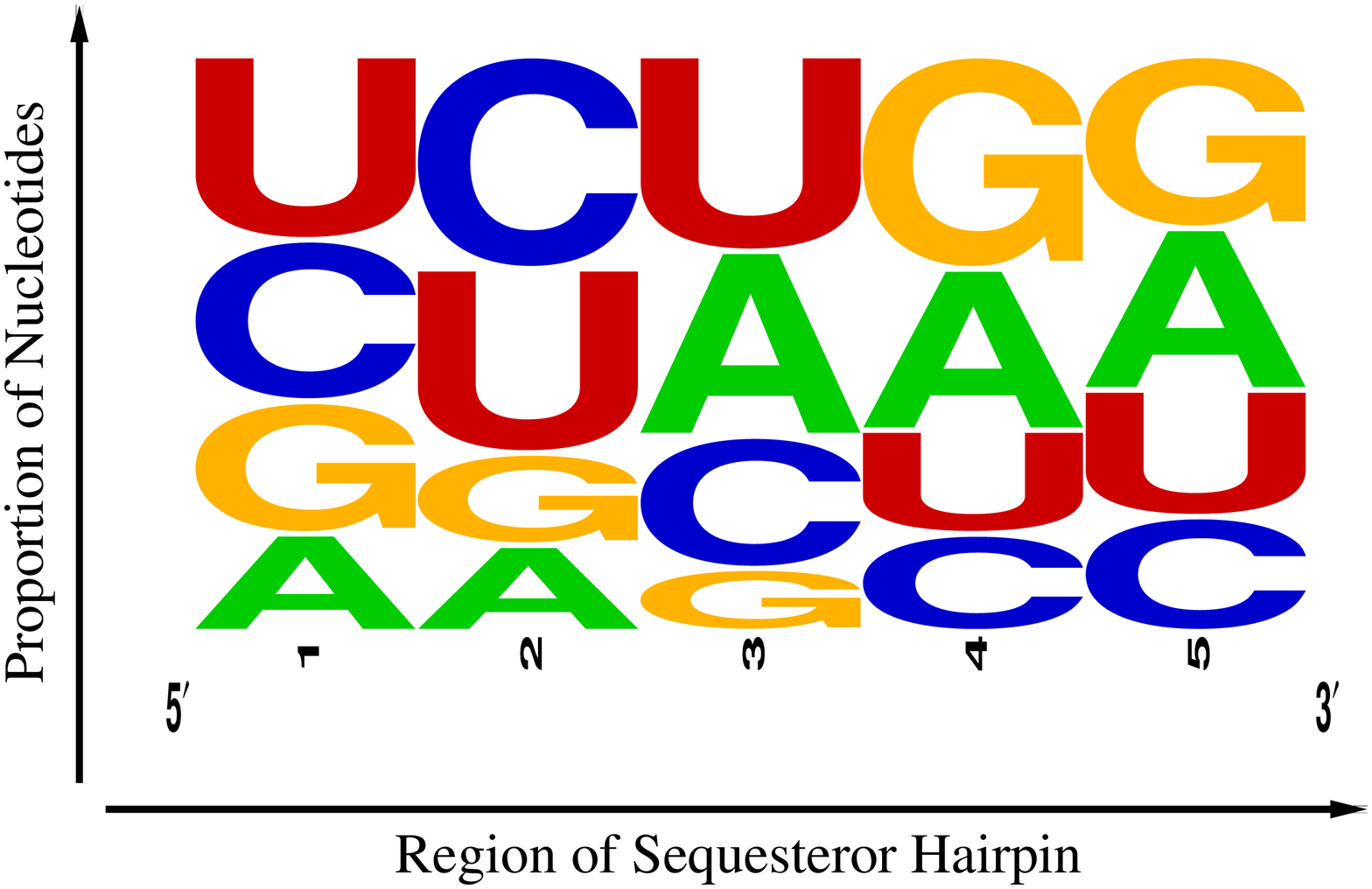}
   \end{center}
\caption{\label{fig:ntdist}
Frequency weighted sequence logos~\cite{Crooks04} for TPP rho-independent transcriptional terminators (left) and Shine-Dalgarno sequesterers (right). Regions 1-5 correspond, respectively, to the first half of the $5^{\prime}$ side of the stem, the second half of the same, the loop, the first half of the $3^{\prime}$ side of the stem, the second half of the same.}
\end{figure}

A second difference lies in the nucleotides frequencies (Table~\ref{tab:stats}) and their distribution among five regions of the hairpins as illustrated in Fig.~\ref{fig:ntdist}.
Here region 3 represents the hairpin loop, with regions 1 and 2 lying along the $5^{\prime}$ side of the hairpin and regions 4 and 5 along the $3^{\prime}$ side.
Terminators exhibit an excess of U in region 5 associated with the beginning of the poly-U pause site, and a weak corresponding enhancement of complementary A nucleotides in region 1.
Sequesterers, in contrast, exhibit an enhancement of A and G in regions 4 and 5 corresponding to the Shine-Dalgarno consensus sequence of AGGAGG, and a corresponding enhancement of complementary C and U in regions 1 and 2.
Another difference is the excess U in the loop region 3 of terminators that can be attributed to an internal pause site allowing time for aptamer and antiterminator folding~\cite{Wickiser05} prior to completion of terminator transcription.
The enhancement of the specifically-binding C and its non-complementary U in regions 1 and 2 of the sequesterer might have been expected to aid in folding efficiency, yet still the terminators, dominated in most regions by the promiscuously-binding U and G, manage to fold with relatively high efficiency.
However, the weak enhancement of specifically-binding A in region 1 of the terminator, complementary to the poly-U pause site in region 5, may play some small role in terminator folding efficiency.

Overall, neither the differences in sequence length nor in nucleotide content appear capable of explaining the difference in folding efficiency between terminators and sequesterers.
The most likely explanation available is simply that the folding efficiencies differ as a result of natural selection.
Selection pressure apparently favors relatively short hairpins and disfavors sequences containing metastable traps in terminators that must fold under the constraint of short transcription time.
This selection pressure is reduced or absent in the case of Shine-Dalgarno sequesterers.
Indeed, as evidenced in Fig.~\ref{fig:efftime}, many sequesterers fail to fold efficiently even on very long time scales.
Perhaps sequesters function in an ensemble of metastable structures, provided the Shine-Dalgarno sequence remains bound, while in contrast transcriptional terminators require very specific structures in order to function~\cite{Wilson95,Lesnik01,Nudler02}.

\section{Specific examples}

\begin{figure}
\begin{center}
\includegraphics[height=0.3\textwidth]{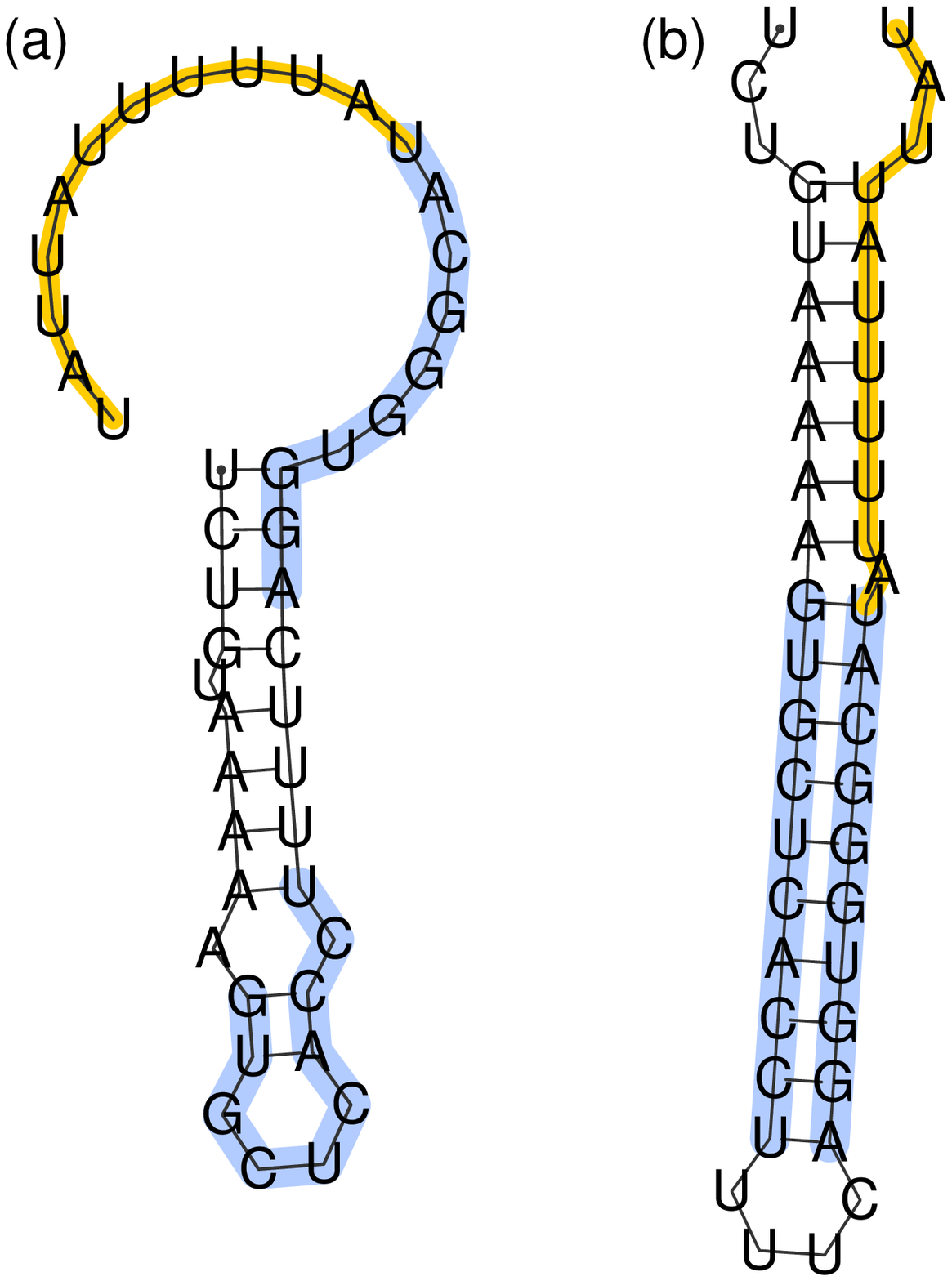}
\includegraphics[height=0.35\textwidth]{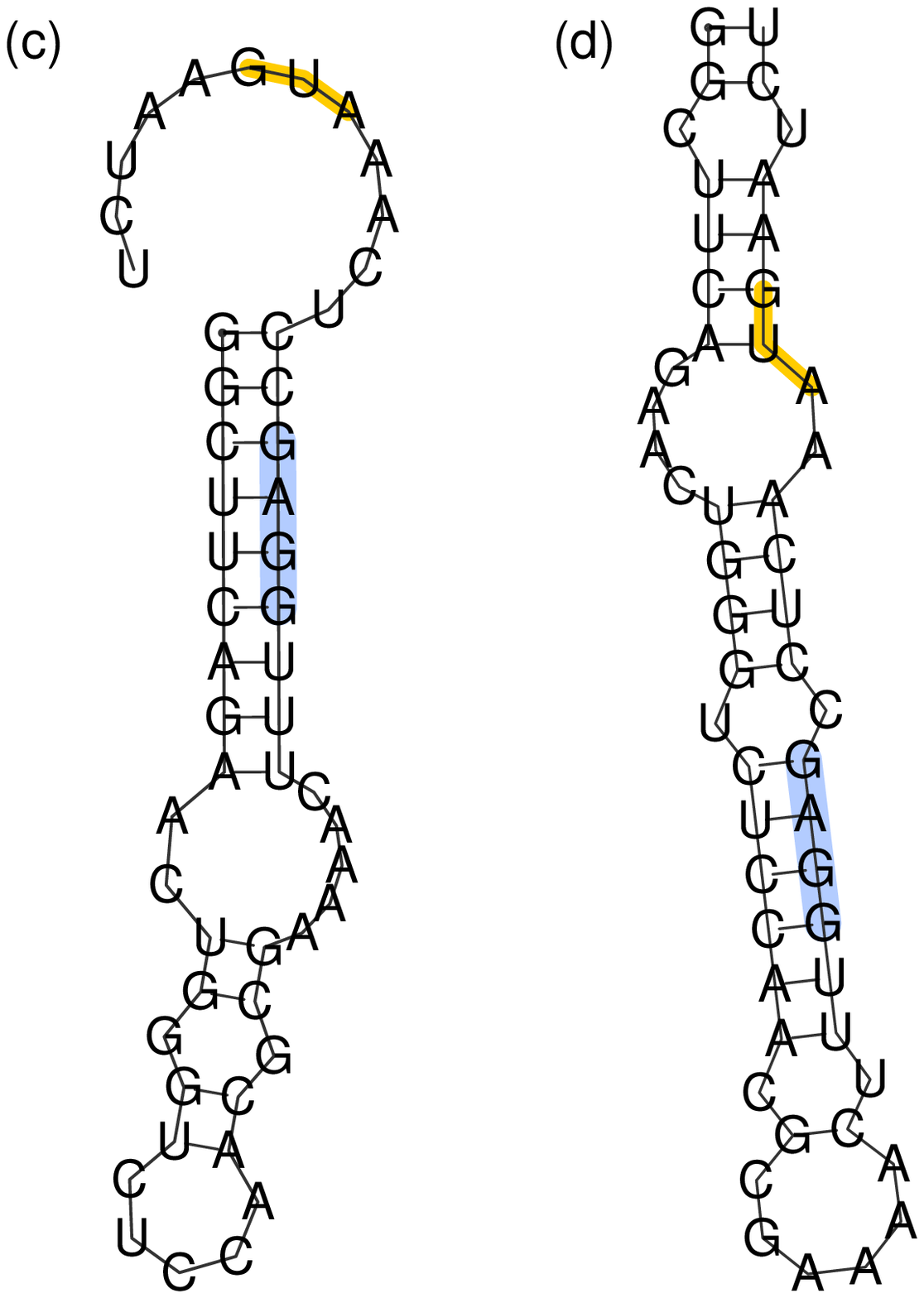}
\end{center}
\caption{\label{fig:folds} Alternate folds of low efficiency terminators and sequesterers. (a,b) Most common specific fold and MFE structure of the {\em Tte-ThiD} terminator sequence. Nucleotides forming stem of terminator are highlighted in blue, while poly-$U$ pause site is in orange.  (c,d) Most common specific fold and MFE structure of {\em Sm-ThiC}. Shine-Dalgarno sequence is highlighted in blue, while the translation start site is highlighted in orange.}
\end{figure}

Here we analyze specific cases of poorly folding terminators and sequesterers.  The most poorly folding terminator is the {\em ThiD} terminator of Thermoanaerobacter tengcongensis ({\em Tte}) which folds with efficiency $e_s$=0.18 at the fastest transcription rate (smallest timescale) $\rho=1/\tau_K R_t$=250 MC steps/nt transcribed.  Similarly, the {\em ThiC} riboswitch of Sinorhizobium meliloti ({\em Sm}) stands out for having the lowest observed efficiency ($e_s=0.292$) at the slowest transcription rate (largest timescale), $\rho$=512000 MC steps/nt transcribed. Two alternate folds of each sequence are illustrated in Fig.~\ref{fig:folds}.  The most common specific fold of {\em Tte-ThiD} (Fig.~\ref{fig:folds}a), which occurs in 35\% of folding attempts, shares no common pairs with the MFE structure (Fig.~\ref{fig:folds}b), which occurs in 6\% of folding attempts.  Likewise, for {\em Sm-ThiC}, the most common specific fold (Fig.~\ref{fig:folds}c) occurs in 10\% of attempts and shares no common pairs with the MFE structure (Fig.~\ref{fig:folds}d), which occurs in 3\% of attempts.

The misfolded terminator (Fig.~\ref{fig:folds}a) lacks the necessary hairpin preceding the poly-$U$ pause site that terminates transcription.  It is notable that the Shine-Dalgarno sequence remains sequestered in the misfolded sequesterer, suggesting that perhaps the function is preserved.  This might explain how low folding efficiency sequesterers could remain functional even while misfolded on the time scale of translation initiation.

\begin{figure}
\begin{center}
\includegraphics[width=0.3\textwidth]{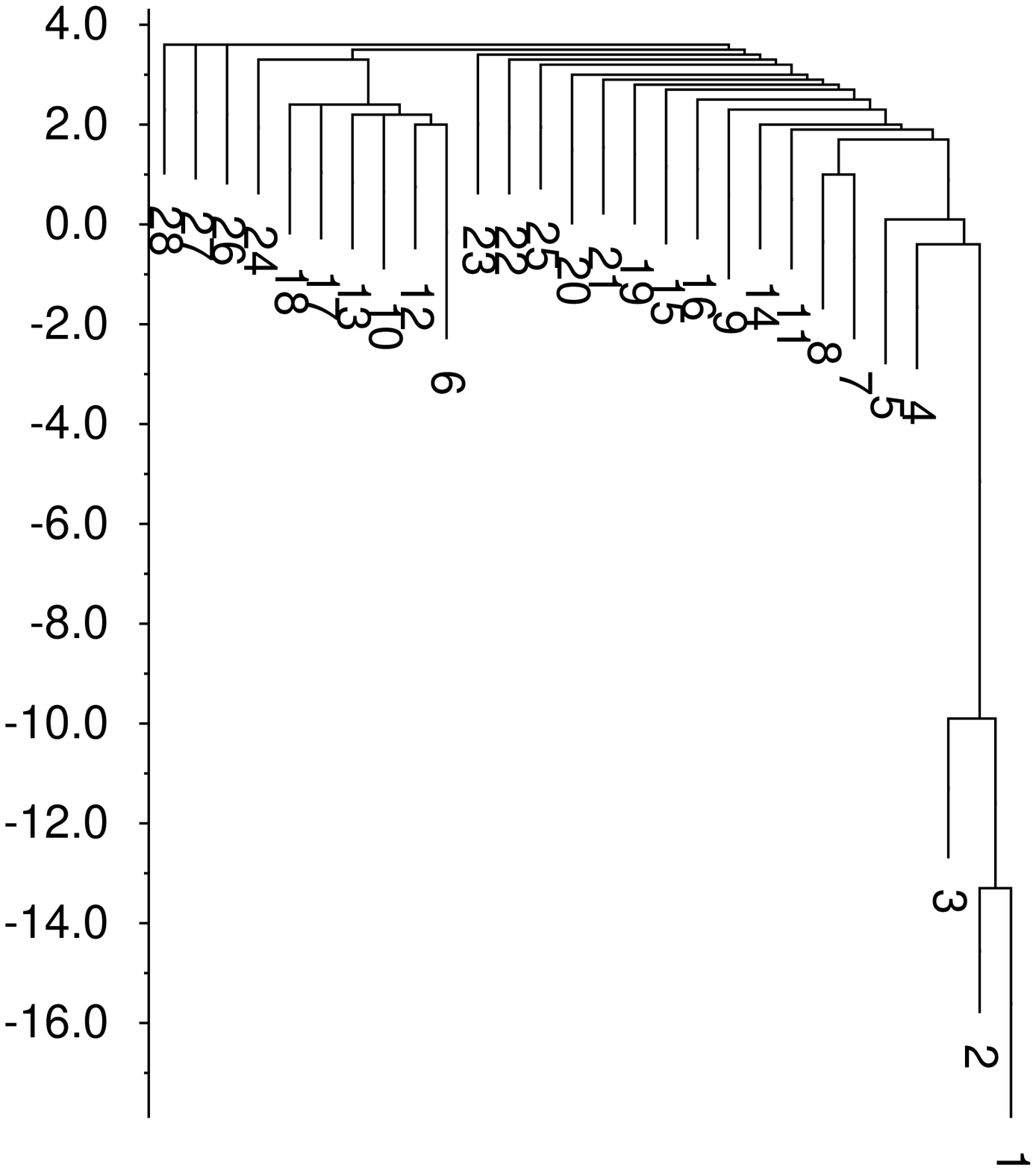}
\includegraphics[width=0.3\textwidth]{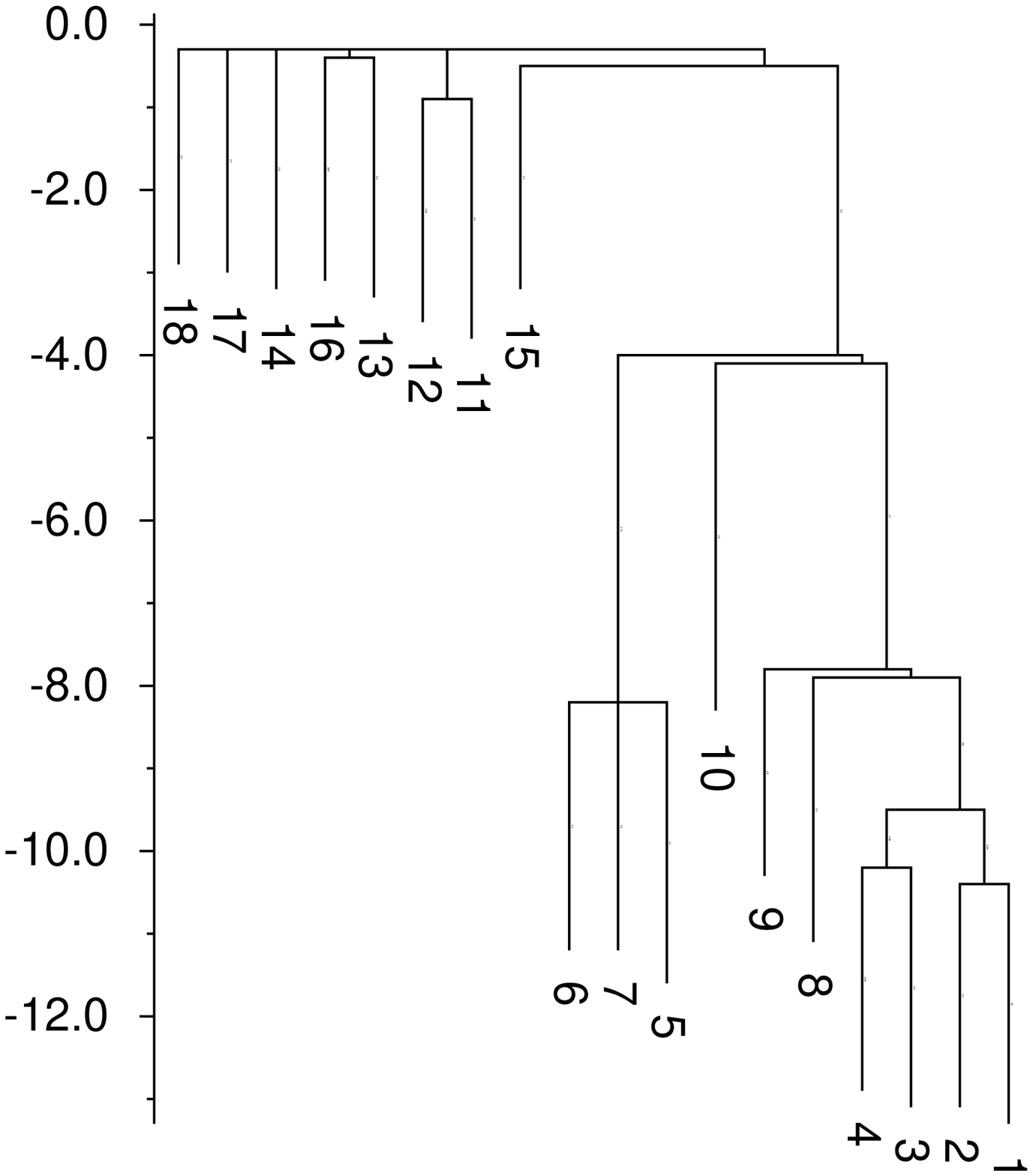}
\end{center}
\caption{\label{fig:tree} Free energy landscapes in units of kcal/mol.  Completely unbound structures have energy 0.  Basins of depth less than 2.5 have been suppressed  {\em Tte-ThiD} (left) structure number 6 corresponds to the most common fold (Fig.~\ref{fig:folds}a).
{\em Sm-ThiC} (right) structure number 5 corresponds to the most common fold (Fig.~\ref{fig:folds}c). In both cases, structure number 1 is the MFE fold (Figs.~\ref{fig:folds}(b,d)).}
\end{figure}

At the largest timescale, $\rho$=512000, the efficiency of {\em Tte-ThiD} rises to 91\%.  To understand the high efficiency of {\em Tte-ThiD} relative to {\em Sm-ThiC} at long times, we compare their free energy landscapes in Fig.~\ref{fig:tree}.
The misfold of {\em Tte-ThiD} is relatively weakly bound (only -2.3 kcal/mol) with a barrier of 5.8 kcal/mol separating the misfold from the MFE structure.  This barrier has high entropy, as it corresponds to complete unfolding followed by almost any single base pairing yielding a net energy for the saddle state of +3.5 kcal/mol.  This high barrier entropy reduces the effective free energy barrier~\cite{Sauerwine11}.  Furthermore, as {\em Tte} is a thermophile, relatively high thermal energy is available to aid in escape from metastable traps.
In contrast, {\em Sm-ThiC} is relatively strongly bound (-11.6 kcal/mol).  The saddle state separating the misfold from the MFE is only partially unbound, at energy -4.0, but the net barrier of 7.6 kcal is nearly 2 kcal/mol (about $3RT$) larger than for {\em Tte-ThiD} and also is relatively low entropy.

The common misfolds of both {\em Tte-ThiD} and {\em Sm-ThiC} share a common feature - their paired nucleotides lie to the $5^{\prime}$ (earlier transcribed) side of the pairs comprising the MFE structures.  That is, they contain structure that can form before the sequence is fully transcribed. To see how widespread this mechanism is, we examined the 16 sequesterers that fold with efficiency less that 0.5 at our standard transcription rate, $\rho$=4000.  In all but one case the most common misfold places the hairpin loop to the $5^{\prime}$ side of its location in the MFE structure. That is, they involve structures that can form earlier in time than the MFE.  The sole exception, is a very short sequence for which a few missing pairs reduce the matched fraction $f$ below 0.7 even while the sequence lies in the basin of the MFE structure.

\section{Conclusions}

In conclusion, this study addressed whether riboswitch transcriptional terminators fold with unusually high efficiency, indicating selection for reliability of folding.  It was shown that transcriptional terminators in TPP riboswitches are unusually easy to fold during transcription in comparison with Shine-Dalgarno sequesterers, resulting in a strongly significant $p$-value for the null hypothesis.  Experimental validation of this prediction might be feasible using optical tweezer studies~\cite{Frieda2012}.

Detailed examination of a specific terminator ({\em Tte-ThiD} and sequesterer ({\em Sm-ThiC}) which fold with relatively low efficiency, reveals a generic mechanism for misfolding, namely trapping into minimum free energy conformations of partially transcribed sequences that become potentially long-lived metastable states of the fully transcribed sequence.  We also suggest that sequesterers may be more tolerant of misfolds than terminators, provided that the Shine-Dalgarno sequence remains bound in the misfolded structure.


\acknowledgements{Acknowledgements}

The authors thank Jay Kadane, Maumita Mandal and Jon Widom for useful discussions.  We acknowledge financial support of this research by the DSF Charitable Foundation.


\conflictofinterests{The authors declare no conflict of interest}

\bibliography{kinetics}
\bibliographystyle{mdpi}

\end{document}